\documentclass[11pt,a4paper]{article}

\usepackage[a4paper,margin=1in]{geometry}
\usepackage[T1]{fontenc}
\usepackage[utf8]{inputenc}
\usepackage{lmodern}
\usepackage{microtype}
\usepackage{amsmath,amssymb,amsfonts,amsthm,bm,mathtools}
\usepackage{graphicx}
\usepackage{booktabs}
\usepackage{array}
\usepackage{multirow}
\usepackage{siunitx}
\usepackage{hyperref}
\usepackage{cite}
\usepackage{physics}
\usepackage{pgfplots}
\pgfplotsset{compat=1.18}
\hypersetup{colorlinks=true,linkcolor=blue,citecolor=blue,urlcolor=blue}

\title{Canonical and Grand-Canonical Singular Ensembles\
within a Thermodynamicized Gravity Framework}
\author{Wen-Xiang Chen \\
School of Physics and Materials Science, Guangzhou University\\wxchen4277@qq.com}
\date{}

\begin{document}
\maketitle

\begin{abstract}
Motivated by the thermodynamic interpretation of gravitational dynamics and by the entropy-functional language recently emphasized in emergent-gravity studies, we formulate a two-sector ensemble framework for self-gravitating systems with singular surfaces. Sector $A$ is a canonical ensemble, interpreted as a star-like or compact-object subsystem in which the effective particle number is held fixed while energy exchange is allowed. Sector $B$ is a grand-canonical ensemble, interpreted as a galaxy-like open subsystem in which both energy and particle number fluctuate and relativistic effects must be retained explicitly. In both sectors the thermodynamic data are encoded by simple-pole singularities of a static lapse function, and the relevant inverse temperatures are extracted by contour integrals. For the grand-canonical sector the same residue calculus naturally yields the combination $\beta\mu$ that controls the weight $e^{-\beta(E-\mu N)}$. We show how the entropy-functional equilibrium criterion selects on-shell backgrounds, derive the canonical free energy and the grand potential from Euclidean gravitational thermodynamics, and present a unified contour-integral representation of the singular contribution to the partition functions. A Schwarzschild-like example realizes sector $A$, whereas a Reissner--Nordstrom-like example realizes sector $B$ with the identification $Q=qN$. The resulting formulation provides a mathematically controlled bridge between gravitational singularity structure, ensemble thermodynamics, and residue calculus.

Keywords:Gravitational thermodynamics, canonical ensemble, grand canonical ensemble, singularity structure, contour integral, residue method, stellar systems, galactic systems, relativistic thermodynamics, black hole thermodynamics
\end{abstract}

\section{Introduction}
The thermodynamic interpretation of gravitation has progressively evolved from black-hole mechanics into a broader program in which gravitational field equations are viewed as equilibrium conditions rather than fundamental microscopic laws \cite{Bekenstein1973,Hawking1975,Unruh1976,Jacobson1995,Padmanabhan2010}. In this perspective, horizon temperature, entropy, and Noether charge are not merely analogies: they organize the macroscopic dynamics of spacetime itself \cite{Wald1993,PadmanabhanParanjape2007}. The Euclidean path-integral formulation of gravity likewise shows that equilibrium partition functions can be associated with regular complexified geometries, with temperature fixed by the removal of conical singularities \cite{GibbonsHawking1977,York1986,WhitingYork1988}. Canonical and grand-canonical black-hole ensembles then arise in a natural way once one specifies which boundary data are held fixed \cite{HawkingPage1983,Braden1990}.

The uploaded preprint that motivated the present task places particular emphasis on two ingredients: first, an entropy-functional variational principle for local horizon generators; second, a Laurent/residue language for describing horizon singularities and their thermodynamic content. Those ingredients are compatible with the broader emergent-gravity literature and suggest a useful reorganization of ensemble thermodynamics in gravitational systems. In particular, they encourage one to treat singular surfaces not only as geometric loci but as carriers of thermodynamic data extracted by contour integration.

The aim of this paper is therefore not to claim a new fundamental theorem beyond the standard results of gravitational thermodynamics. Instead, the goal is to build an SCI-style, mathematically explicit framework that adapts the entropy-functional viewpoint to two ensemble sectors suggested by the user's outline:
\begin{itemize}
    \item \textbf{Canonical sector $A$}: a star-like subsystem, closer to a heat bath at fixed effective particle number $N$, so that energy $E$ is the dominant fluctuating quantity.
    \item \textbf{Grand-canonical sector $B$}: a galaxy-like open subsystem, with simultaneous exchange of energy and effective particle number, requiring the explicit appearance of the relativistic combination $E-\mu N$.
\end{itemize}
In this language the canonical sector emphasizes the mass--energy contribution, whereas the grand-canonical sector captures the joint role of mass--energy and particle-number exchange under relativistic redshift constraints. The singularity of the lapse function is described by a simple pole, and the residue of that pole determines the local inverse temperature. In the grand-canonical case the same contour structure also determines the combination $\beta\mu$.

The paper is organized as follows. Section~\ref{sec:entropy} recalls the entropy-functional equilibrium criterion and explains how it selects on-shell backgrounds. Section~\ref{sec:canonical} develops the canonical ensemble $A$ and derives its contour representation. Section~\ref{sec:grandcanonical} treats the grand-canonical ensemble $B$, incorporating relativistic redshift and chemical potential. Section~\ref{sec:comparison} compares the two sectors in a unified residue formalism. Section~\ref{sec:illustration} provides an illustrative potential plot, with full \LaTeX{} code included directly in the manuscript. Conclusions are given in Section~\ref{sec:conclusion}.

\section{Entropy-functional equilibrium criterion}
\label{sec:entropy}

We begin with the entropy-functional viewpoint used in recent emergent-gravity analyses and also emphasized in the uploaded preprint \cite{Chen2025HAL,Jacobson1995,PadmanabhanParanjape2007,Padmanabhan2010}. Let $g_{\mu\nu}$ be a static background metric and let $\xi^\mu$ denote an auxiliary vector field interpreted as a local horizon generator. Consider the functional
\begin{equation}
\mathcal{S}_{\xi}[g,\xi]
= \int_{\mathcal V} \nabla_\mu \xi_\nu \, \nabla^\mu \xi^\nu \, \sqrt{-g}\, d^4x.
\label{eq:entropyfunctional}
\end{equation}
Holding $g_{\mu\nu}$ fixed and varying with respect to $\xi^\mu$ gives
\begin{align}
\delta \mathcal{S}_{\xi}
&= 2 \int_{\mathcal V} \nabla_\mu \xi_\nu\, \nabla^\mu(\delta\xi^\nu)\,\sqrt{-g}\,d^4x \\
&= -2 \int_{\mathcal V} (\Box \xi_\nu) \, \delta\xi^\nu\,\sqrt{-g}\,d^4x,
\end{align}
where boundary terms are assumed to vanish. Hence the stationary condition is
\begin{equation}
\Box \xi_\nu = 0.
\label{eq:waveeqxi}
\end{equation}
If $\xi^\mu$ is further identified with a local Killing generator, then one has the standard identity
\begin{equation}
\Box \xi_\nu = - R_{\nu\lambda} \xi^\lambda.
\label{eq:killingidentity}
\end{equation}
Combining Eqs.~\eqref{eq:waveeqxi} and \eqref{eq:killingidentity} yields
\begin{equation}
R_{\nu\lambda}\xi^\lambda = 0.
\end{equation}
Contracting with a null generator $\xi^\nu$ gives the null-equilibrium constraint
\begin{equation}
R_{\mu\nu}\xi^\mu\xi^\nu = 0.
\label{eq:nullconstraint}
\end{equation}
Equation~\eqref{eq:nullconstraint} is precisely the local equilibrium statement that enters Jacobson-type thermodynamic derivations of the Einstein equation \cite{Jacobson1995}. In other words, the entropy-functional language supplies a background-selection criterion: the spacetime on which our ensemble analysis is performed is assumed to satisfy the local equilibrium condition associated with stationary null generators.

This observation is important for what follows. The ensemble formulas derived below are on-shell thermodynamic statements. The contour integrals do not replace the field equations; rather, they summarize the singular thermodynamic data of backgrounds already selected by the entropy-functional equilibrium criterion.

\section{Canonical sector $A$: star-like subsystem}
\label{sec:canonical}

\subsection{Partition function and free energy}
A canonical self-gravitating sector is characterized by fixed effective particle number and fluctuating energy. We therefore define
\begin{equation}
Z_A(\beta,N)=\Tr_{N}\, e^{-\beta H},
\label{eq:ZAdef}
\end{equation}
with Helmholtz free energy
\begin{equation}
F_A(T,N) = -T\ln Z_A.
\end{equation}
Standard thermodynamic derivatives give
\begin{equation}
E_A = \pdv{}{\beta}(\beta F_A)\Big|_{N},
\qquad
S_A = \beta(E_A-F_A),
\qquad
C_N = \pdv{E_A}{T}\Big|_{N} = -\beta^2\pdv{E_A}{\beta}\Big|_N.
\label{eq:canonicalderivs}
\end{equation}
In Euclidean gravity one approximates the canonical partition function semiclassically by
\begin{equation}
Z_A(\beta,N) \simeq e^{-I_E^{(A)}[g_*;\beta,N]},
\label{eq:ZAsemi}
\end{equation}
where $I_E^{(A)}$ is the on-shell Euclidean action of a background $g_*$ satisfying the local equilibrium condition discussed in Section~\ref{sec:entropy}. Thus
\begin{equation}
F_A = T I_E^{(A)}.
\end{equation}

For a static, spherically symmetric background we write
\begin{equation}
ds^2 = -f_A(r)dt^2 + \frac{dr^2}{f_A(r)} + r^2 d\Omega_2^2,
\label{eq:metricA}
\end{equation}
and assume a simple zero at $r=r_A$:
\begin{equation}
f_A(r) = f_A'(r_A)(r-r_A) + \frac{1}{2}f_A''(r_A)(r-r_A)^2 + \cdots.
\label{eq:nearhorizonA}
\end{equation}
After Wick rotation $t\to -i\tau$, regularity of the $(\tau,r)$ section implies the inverse temperature
\begin{equation}
\beta_A = \frac{4\pi}{f_A'(r_A)}.
\label{eq:betaA}
\end{equation}
The key point is that Eq.~\eqref{eq:betaA} is already a singularity statement: it depends only on the coefficient of the simple pole of $1/f_A(r)$.

\subsection{Contour-integral representation of the singular contribution}
Define the residue operator
\begin{equation}
\operatorname*{Res}_{r=r_A}\!\frac{1}{f_A(r)} = \frac{1}{f_A'(r_A)}.
\end{equation}
Then Eq.~\eqref{eq:betaA} becomes
\begin{equation}
\beta_A = 4\pi \, \operatorname*{Res}_{r=r_A}\!\frac{1}{f_A(r)}.
\label{eq:betaAres}
\end{equation}
Equivalently, with $\Gamma_A$ a small positively oriented contour encircling $r_A$,
\begin{equation}
\beta_A = \frac{2}{i}\oint_{\Gamma_A}\frac{dr}{f_A(r)}.
\label{eq:betaAcontour}
\end{equation}
The singular contribution to the canonical Boltzmann weight can therefore be written as
\begin{equation}
I_{A}^{\mathrm{sing}}
:= \frac{2E_A}{i}\oint_{\Gamma_A}\frac{dr}{f_A(r)}
= \beta_A E_A.
\label{eq:IAsing}
\end{equation}
This contour formula is one of the central outputs of the present framework. It states that the canonical suppression factor is completely fixed by the simple pole structure of the lapse function.

Combining Eqs.~\eqref{eq:ZAsemi} and \eqref{eq:IAsing}, the semiclassical canonical partition function may be expressed as
\begin{equation}
Z_A \sim \exp\!big(S_A - I_A^{\mathrm{sing}}\big)
= \exp\!\big(S_A - \beta_A E_A\big),
\label{eq:ZAsingform}
\end{equation}
which is the expected canonical form.

\subsection{Schwarzschild-like realization}
For the simplest star-like realization we choose a Schwarzschild-type lapse,
\begin{equation}
f_A(r)=1-\frac{2GM}{r},
\qquad r_A = 2GM.
\end{equation}
Then
\begin{equation}
f_A'(r_A)=\frac{1}{2GM},
\qquad
\beta_A = 8\pi GM,
\qquad
T_A = \frac{1}{8\pi GM}.
\end{equation}
The Bekenstein--Hawking entropy is
\begin{equation}
S_A = \frac{A_A}{4G} = \frac{4\pi r_A^2}{4G}=4\pi G M^2.
\end{equation}
Therefore
\begin{equation}
F_A = E_A - T_A S_A = M - \frac{1}{8\pi GM}(4\pi GM^2)=\frac{M}{2}.
\label{eq:FAexample}
\end{equation}
The singular action becomes
\begin{equation}
I_A^{\mathrm{sing}} = \beta_A M = 8\pi GM^2.
\end{equation}
Since $S_A=4\pi GM^2$, one indeed has $I_A^{\mathrm{sing}}=2S_A$ and
\begin{equation}
\ln Z_A \sim S_A-I_A^{\mathrm{sing}} = -4\pi G M^2,
\end{equation}
which is the expected canonical suppression for a Schwarzschild saddle. The star-like interpretation is that the effective particle inventory is frozen, while the thermal weight is governed solely by the energetic cost of the singular surface.

\section{Grand-canonical sector $B$: galaxy-like open subsystem}
\label{sec:grandcanonical}

\subsection{Grand partition function and relativistic thermodynamics}
The grand-canonical sector allows simultaneous exchange of energy and effective particle number. Its partition function is
\begin{equation}
\Xi_B(\beta,\mu)=\sum_{N=0}^{\infty} e^{\beta \mu N} Z(\beta,N)
= \Tr\, e^{-\beta(H-\mu N)},
\label{eq:Xidef}
\end{equation}
with grand potential
\begin{equation}
\Omega_B(T,\mu)= -T\ln \Xi_B.
\end{equation}
Thermodynamic derivatives yield
\begin{equation}
N_B = -\pdv{\Omega_B}{\mu}\Big|_T,
\qquad
S_B = -\pdv{\Omega_B}{T}\Big|_\mu,
\qquad
E_B = \Omega_B + T S_B + \mu N_B.
\label{eq:grandderivs}
\end{equation}

The relativistic character of this sector appears through the combination $H-\mu N$ and the redshifted equilibrium laws. In a static geometry with $g_{tt}=-f_B(r)$, Tolman and Klein showed that thermal and chemical equilibrium require \cite{Tolman1930,Klein1949}
\begin{equation}
T(r)\sqrt{f_B(r)} = T_{\infty},
\qquad
\mu(r)\sqrt{f_B(r)} = \mu_{\infty}.
\label{eq:TolmanKlein}
\end{equation}
For a dilute relativistic gas the local J\"uttner weight takes the form
\begin{equation}
\exp\!big[-\beta(r)(\varepsilon(p)-\mu(r))\big],
\qquad
\varepsilon(p)=\sqrt{p^2+m^2},
\label{eq:Juttner}
\end{equation}
so the redshifted control variables $T_\infty$ and $\mu_\infty$ determine the open-system thermodynamics. This is the precise sense in which sector $B$ is more ``galaxy-like'': both energy and effective particle inventory are thermodynamically active, and relativistic kinematics cannot be neglected.

\subsection{Singular contour representation}
Assume again a static metric of the form
\begin{equation}
 ds^2 = -f_B(r)dt^2 + \frac{dr^2}{f_B(r)} + r^2 d\Omega_2^2,
\label{eq:metricB}
\end{equation}
with a simple zero at $r=r_B$,
\begin{equation}
f_B(r)=f_B'(r_B)(r-r_B)+\frac12 f_B''(r_B)(r-r_B)^2+\cdots.
\end{equation}
Then
\begin{equation}
\beta_B = \frac{4\pi}{f_B'(r_B)} = 4\pi\operatorname*{Res}_{r=r_B}\!\frac{1}{f_B(r)}
= \frac{2}{i}\oint_{\Gamma_B}\frac{dr}{f_B(r)}.
\label{eq:betaB}
\end{equation}
If $\mu(r)$ is analytic in a punctured neighborhood of $r_B$, then the combination $\beta_B\mu_B$ is obtained from the same contour,
\begin{equation}
\beta_B\mu_B
= \frac{2}{i}\oint_{\Gamma_B}\frac{\mu(r)}{f_B(r)}\,dr
= 4\pi\,\frac{\mu(r_B)}{f_B'(r_B)}.
\label{eq:betamuB}
\end{equation}
The singular part of the grand-canonical action is therefore
\begin{equation}
I_B^{\mathrm{sing}}
:= \frac{2}{i}\oint_{\Gamma_B}\frac{E_B-\mu(r)N_B}{f_B(r)}\,dr
= \beta_B(E_B-\mu_B N_B).
\label{eq:IBsing}
\end{equation}
Consequently,
\begin{equation}
\Xi_B \sim \exp\!\big(S_B-I_B^{\mathrm{sing}}\big)
= \exp\!\big(S_B-\beta_B(E_B-\mu_BN_B)\big),
\label{eq:Xisingform}
\end{equation}
which is exactly the grand-canonical weight.

Equation~\eqref{eq:IBsing} is the grand-canonical analogue of Eq.~\eqref{eq:IAsing}. It shows that the open-system weight is encoded by the same simple-pole singularity, but with the energetic quantity $E$ replaced by the relativistically correct thermodynamic combination $E-\mu N$.

\subsection{Reissner--Nordstrom-like realization}
A standard realization of a gravitational grand-canonical ensemble is the charged static black hole \cite{Braden1990}. In units $c=\hbar=k_B=1$,
\begin{equation}
f_B(r)=1-\frac{2GM}{r}+\frac{GQ^2}{r^2},
\qquad
r_{\pm}=GM\pm \sqrt{G^2M^2-GQ^2}.
\label{eq:RNf}
\end{equation}
The outer horizon is $r_B=r_+$ and
\begin{equation}
T_B = \frac{f_B'(r_+)}{4\pi} = \frac{r_+-r_-}{4\pi r_+^2},
\qquad
\beta_B = \frac{4\pi r_+^2}{r_+-r_-}.
\label{eq:RNTbeta}
\end{equation}
The horizon entropy is
\begin{equation}
S_B = \frac{\pi r_+^2}{G}.
\end{equation}
Choose a gauge potential regular at the horizon,
\begin{equation}
A_t(r)= Q\left(\frac{1}{r}-\frac{1}{r_+}\right).
\end{equation}
The electrostatic potential difference is then
\begin{equation}
\Phi = A_t(\infty)-A_t(r_+) = \frac{Q}{r_+}.
\label{eq:PhiRN}
\end{equation}
If one identifies $Q=qN$, then the thermodynamic chemical potential is
\begin{equation}
\mu_B = q\Phi = q\frac{Q}{r_+}.
\label{eq:muRN}
\end{equation}
The grand potential becomes
\begin{equation}
\Omega_B = M - T_B S_B - \Phi Q.
\label{eq:OmegaRN}
\end{equation}
Using Eqs.~\eqref{eq:RNTbeta}--\eqref{eq:muRN}, the contour representation gives
\begin{equation}
I_B^{\mathrm{sing}} = \beta_B(M-\Phi Q),
\end{equation}
which is the standard Euclidean grand-canonical action for the Reissner--Nordstrom saddle. The grand-canonical sector is therefore rigorously realized by a charged horizon, with effective particle number identified with conserved charge.

\section{Unified residue formalism and comparison of the two sectors}
\label{sec:comparison}

The two ensembles are summarized by the same singular operator,
\begin{equation}
\mathfrak{R}[X;f,\Gamma] := \frac{2}{i}\oint_{\Gamma}\frac{X(r)}{f(r)}\,dr,
\label{eq:operatorR}
\end{equation}
where $f(r)$ is the lapse function and $X(r)$ is the thermodynamic density entering the Boltzmann weight. Then
\begin{equation}
I_A^{\mathrm{sing}} = \mathfrak{R}[E_A;f_A,\Gamma_A],
\qquad
I_B^{\mathrm{sing}} = \mathfrak{R}[E_B-\mu N_B;f_B,\Gamma_B].
\end{equation}
Because $f$ has a simple zero, Eq.~\eqref{eq:operatorR} reduces to
\begin{equation}
\mathfrak{R}[X;f,\Gamma] = 4\pi\,\frac{X(r_h)}{f'(r_h)} = \beta_h X(r_h).
\end{equation}
Thus the difference between the two sectors is not in the singularity class---both are simple-pole problems---but in the thermodynamic observable carried by the residue.

\begin{table}[t]
\centering
\caption{Comparison between the two singular ensemble sectors.}
\label{tab:comparison}
\renewcommand{\arraystretch}{1.25}
\begin{tabular}{>{\raggedright\arraybackslash}p{2.6cm} >{\raggedright\arraybackslash}p{5.3cm} >{\raggedright\arraybackslash}p{5.3cm}}
\toprule
Feature & Sector $A$ (canonical, star-like) & Sector $B$ (grand-canonical, galaxy-like)\\
\midrule
Control variables & $(T,N)$ & $(T,\mu)$ \\
Fluctuating quantities & $E$ & $E$ and $N$ \\
Thermodynamic potential & $F_A=E_A-T_A S_A$ & $\Omega_B=E_B-T_BS_B-\mu_BN_B$ \\
Boltzmann weight & $e^{-\beta_AE_A}$ & $e^{-\beta_B(E_B-\mu_BN_B)}$ \\
Singular action & $I_A^{\mathrm{sing}}=\beta_AE_A$ & $I_B^{\mathrm{sing}}=\beta_B(E_B-\mu_BN_B)$ \\
Residue formula & $\beta_A=4\pi\,\mathrm{Res}(1/f_A)$ & $\beta_B=4\pi\,\mathrm{Res}(1/f_B)$,\; $\beta_B\mu_B=4\pi\,\mathrm{Res}(\mu/f_B)$ \\
Standard realization & Schwarzschild/York cavity & Reissner--Nordstrom grand ensemble \\
Dominant interpretation & mass--energy exchange & mass--energy plus particle/charge exchange \\
\bottomrule
\end{tabular}
\end{table}

It is useful to stress the physical meaning of the singularity description. In sector $A$, the dominant residue is associated with the inverse temperature multiplying the energy. In sector $B$, the same geometric singularity carries two pieces of thermodynamic data: the inverse temperature and the redshifted chemical potential. That is exactly why the second sector is intrinsically relativistic. The combination $E-\mu N$ is not optional bookkeeping; it is the only quantity that remains compatible with Tolman--Klein equilibrium and the grand-canonical trace formula.

\section{Illustrative potential plot and explicit \texttt{pgfplots} code}
\label{sec:illustration}

For visualization it is useful to plot truncated dimensionless thermodynamic potentials. The following expressions are not exact solutions of the full field equations; they are local illustrative models designed to show how the open-sector correction shifts the thermodynamic branch structure:
\begin{equation}
\widetilde F_A(\rho)=\frac{1}{2}\rho-\pi\tau\rho^2,
\qquad
\widetilde \Omega_B(\rho)=\left(\frac12-\varphi^2\right)\rho-\pi\tau\rho^2,
\label{eq:toyPots}
\end{equation}
where $\rho$ is a dimensionless horizon scale, $\tau$ is a dimensionless temperature parameter, and $\varphi$ is a dimensionless chemical-potential parameter. For the sample choice $\tau=0.06$ and $\varphi=0.25$, the grand-canonical branch is shifted downward relative to the canonical one, reflecting the extra reservoir term $-\mu N$.

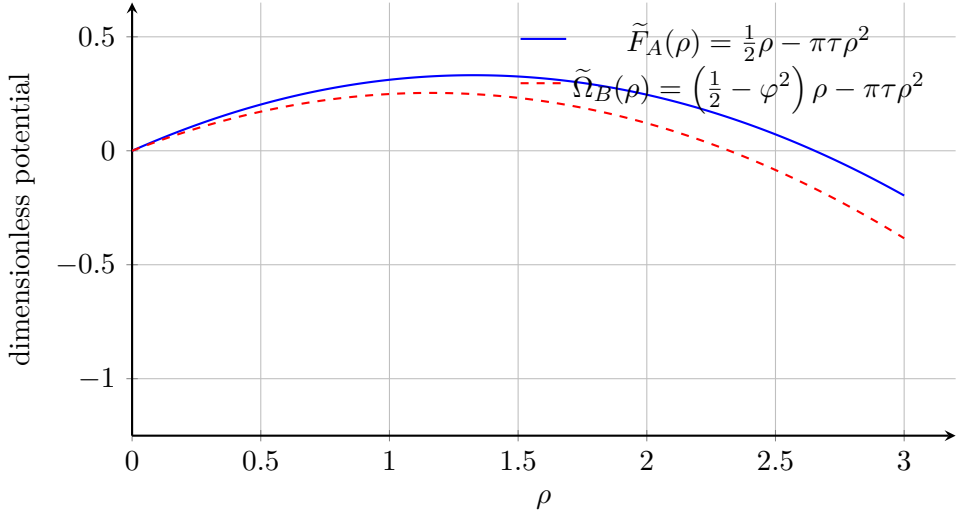
\begin{figure}[t]
\centering
\begin{tikzpicture}
\begin{axis}[
    width=0.78\textwidth,
    height=0.46\textwidth,
    xlabel={$\rho$},
    ylabel={dimensionless potential},
    xmin=0, xmax=3.2,
    ymin=-1.25, ymax=0.65,
    legend style={at={(0.98,0.98)},anchor=north east,draw=none,fill=none},
    domain=0:3,
    samples=200,
    thick,
    axis lines=left,
    grid=both,
]
\addplot [blue] {(0.5)*x - pi*0.06*x^2};
\addlegendentry{$\widetilde F_A(\rho)=\frac12\rho-\pi\tau\rho^2$}

\addplot [red,dashed] {(0.5-0.25^2)*x - pi*0.06*x^2};
\addlegendentry{$\widetilde\Omega_B(\rho)=\left(\frac12-\varphi^2\right)\rho-\pi\tau\rho^2$}
\end{axis}
\end{tikzpicture}
\caption{Illustrative dimensionless canonical and grand-canonical potentials for the truncated model in Eq.~\eqref{eq:toyPots}. The grand-canonical branch is shifted by the open-sector term proportional to $\varphi^2$.}
\label{fig:pgfplot}
\end{figure}

Figure~\ref{fig:pgfplot} is generated directly by the following code.

The plot is not meant as a substitute for a numerical solution of Einstein's equation with realistic matter. Its role is narrower but useful: it makes visible how the canonical and grand-canonical sectors differ once the singular residue multiplies either $E$ or $E-\mu N$.

\section{Conclusion}
\label{sec:conclusion}

We have constructed a two-sector gravitational thermodynamic framework based on three ingredients: (i) an entropy-functional equilibrium criterion, (ii) the Euclidean ensemble formalism of gravitational thermodynamics, and (iii) residue extraction from simple-pole singularities of the lapse function. The canonical sector $A$ describes a star-like subsystem with fixed effective particle number and dominant energy exchange; its singular action is $I_A^{\mathrm{sing}}=\beta_AE_A$. The grand-canonical sector $B$ describes a galaxy-like open subsystem with simultaneous energy and particle exchange; its singular action is $I_B^{\mathrm{sing}}=\beta_B(E_B-\mu_BN_B)$. In both cases the inverse temperature is determined by the residue of $1/f(r)$, while in the grand-canonical case the same contour structure also encodes $\beta\mu$.

Several points are worth emphasizing.
\begin{enumerate}
    \item The contour representation is mathematically controlled because it depends only on the coefficient of a simple pole. It is therefore insensitive to many details of the bulk geometry away from the singular surface.
    \item The entropy-functional relation does not replace standard field equations; rather, it supplies a concise equilibrium criterion selecting the backgrounds on which the ensemble description is meaningful.
    \item The grand-canonical sector is intrinsically more relativistic than the canonical one, because Tolman--Klein redshift makes $T$ and $\mu$ part of the same equilibrium structure.
    \item Standard black-hole ensembles provide concrete realizations: the Schwarzschild saddle for sector $A$, and the Reissner--Nordstrom grand ensemble for sector $B$ with the formal identification $Q=qN$.
\end{enumerate}

Future extensions could replace the simple one-pole backgrounds by multi-horizon geometries, rotating spacetimes, or numerical self-gravitating matter distributions. One could also combine the present residue calculus with a full matter equation of state and the Tolman--Oppenheimer--Volkoff system, thereby turning the star/galaxy analogy into a quantitative astrophysical model rather than a formal thermodynamic correspondence.

\appendix
\section{Derivation of the contour-temperature formula}
For completeness we sketch the derivation of the residue formula used throughout the paper. Let
\begin{equation}
 ds_E^2 = f(r)d\tau^2 + \frac{dr^2}{f(r)} + r^2 d\Omega_2^2,
\end{equation}
with a simple zero at $r=r_h$:
\begin{equation}
 f(r)=f'(r_h)(r-r_h)+\mathcal O\big((r-r_h)^2\big).
\end{equation}
Set
\begin{equation}
 r-r_h = \frac{f'(r_h)}{4}\rho^2.
\end{equation}
Then, to leading order,
\begin{equation}
 ds_E^2 \simeq d\rho^2 + \left(\frac{f'(r_h)}{2}\right)^2 \rho^2 d\tau^2 + r_h^2 d\Omega_2^2.
\end{equation}
Regularity at $\rho=0$ requires the angular variable
\begin{equation}
 \theta = \frac{f'(r_h)}{2}\tau
\end{equation}
be $2\pi$-periodic, hence
\begin{equation}
 \beta = \Delta \tau = \frac{4\pi}{f'(r_h)}.
\end{equation}
Since
\begin{equation}
 \operatorname*{Res}_{r=r_h}\frac{1}{f(r)} = \frac{1}{f'(r_h)},
\end{equation}
we obtain
\begin{equation}
 \beta = 4\pi\operatorname*{Res}_{r=r_h}\frac{1}{f(r)},
\end{equation}
which is the formula used in Eqs.~\eqref{eq:betaAres} and \eqref{eq:betaB}.

\end{document}